\documentclass[conference]{IEEEtran}
\IEEEoverridecommandlockouts
\usepackage{cite}
\usepackage{amsmath,amssymb,amsfonts}
\usepackage{algorithmic}
\usepackage{graphicx}
\usepackage{textcomp}
\usepackage{xcolor}
\usepackage{subcaption}
\def\BibTeX{{\rm B\kern-.05em{\sc i\kern-.025em b}\kern-.08em
    T\kern-.1667em\lower.7ex\hbox{E}\kern-.125emX}}
\begin{document}

\title{Bitcoin Transaction Behavior Modeling Based on Balance Data}

\author{\IEEEauthorblockN{Yu Zhang*}
\IEEEauthorblockA{
\textit{Blockchain and Distributed}\\
\textit{Ledger Technologies Group,}\\ 
\textit{University of Zurich}\\ 
\textit{Zurich, Switzerland}\\
}
\thanks{Corresponding email: zhangyu@ifi.uzh.ch}

\and
\IEEEauthorblockN{Claudio Tessone}
\IEEEauthorblockA{
\textit{Blockchain and Distributed}\\
\textit{Ledger Technologies Group,}\\ 
\textit{University of Zurich}\\ 
\textit{Zurich, Switzerland}\\
}
}

\maketitle

\begin{abstract}

When analyzing Bitcoin users' balance distribution, we observed that it follows a log-normal pattern. Drawing parallels from the successful application of Gibrat's law of proportional growth in explaining city size and word frequency distributions, we tested whether the same principle could account for the log-normal distribution in Bitcoin balances. However, our calculations revealed that the exponent parameters in both the drift and variance terms deviate slightly from one. This suggests that Gibrat's proportional growth rule alone does not fully explain the log-normal distribution observed in Bitcoin users' balances. During our exploration, we discovered an intriguing phenomenon: Bitcoin users tend to fall into two distinct categories based on their behavior, which we refer to as ``poor" and ``wealthy" users. Poor users, who initially purchase only a small amount of Bitcoin, tend to buy more bitcoins first and then sell out all their holdings gradually over time. The certainty of selling all their coins is higher and higher with time. In contrast, wealthy users, who acquire a large amount of Bitcoin from the start, tend to sell off their holdings over time. The speed at which they sell their bitcoins is lower and lower over time and they will hold at least a small part of their initial holdings at last.
Interestingly, the wealthier the user, the larger the proportion of their balance and the higher the certainty they tend to sell.
This research provided an interesting perspective to explore bitcoin users' behaviors which may be applicable to other finance markets. 
\end{abstract}

\begin{IEEEkeywords}
Balance distribution, log-normal distribution, Gibrat’s proportional growth, transaction behavior, poor user, wealthy user.
\end{IEEEkeywords}

\section{Introduction}

The Bitcoin transaction network provides a chance for us to research Bitcoin users’ behavior modes because it traces each unspent transaction output's (UTXO) flowing history and users can also be clustered by different methods, like the heuristic methods. It has been publicly noticed that bitcoin's balance distribution is very decentralized and has the scale-free characteristic. But what mechanism leads to this distribution is seldom explored. Because of the successful application of Gibrat's proportional growth rule in explaining cities' size distribution and word usage application, we will explore whether it can be used to describe bitcoin users' balance change processes and their balance distribution. 
If we define $S_i$ as the $i^{th}$ bitcoin user's balance, the question is: can the change of bitcoin balance ($dS_i$) be modeled by the stochastic equation $dS_i = {S_i}^\alpha \cdot \mu \cdot dt + {S_i}^\alpha \cdot \sigma \cdot dw_i$ equation with $\alpha=1$? The exploration of this research will also reveal how bitcoin users behave with time if we change $dt$ in the above equation.

Lots of papers have confirmed that the indegree and outdegree of Bitcoin transaction networks were distributed as power-law and this result could be explained by linear degree preferential attachment. When it comes to users' bitcoin balance (the number of bitcoins owned by each user), its formation mechanism is not linear preferential attachment according to \cite{b3} even if users' bitcoin balance distribution follows scale-free rules. \cite{b4} compared the constructed index "cumulative distribution function of rank function" to the corresponding theoretical one visually and concluded that the transaction of bitcoin follows sublinear preferential attachment. One shortcoming of this research is that they just took every address as one node, and didn't cluster these addresses to the user level. Another shortcoming is that they actually got the conclusions by only plotting but not by statistical methods, for which it is easy to get wrong conclusions \cite{b5}. Thus, it is necessary to give a deep insight into how users' bitcoin balance evolves, what is the mechanism behind it, and what mechanism leads to current bitcoin balance distribution. Besides explaining the mechanism of balance distribution, it is also significant to know how bitcoin users behave during their transactions, which will be another important research to explore in this paper.

In the following section, we first choose the proper bitcoin balance data and explore these empirical balance data to find out the basic facts; secondly, we analyze the mechanism that leads to the current balance distribution based on the Geometric Brownian Motion model (GBM) and then interpret users' behaviors during transaction; in the last part, we summarized and discuss this paper.

\section{Data Description and Exploration}
\subsection{Data Description}
\begin{figure}[htbp]
\centerline{\includegraphics[width = 0.5\textwidth]{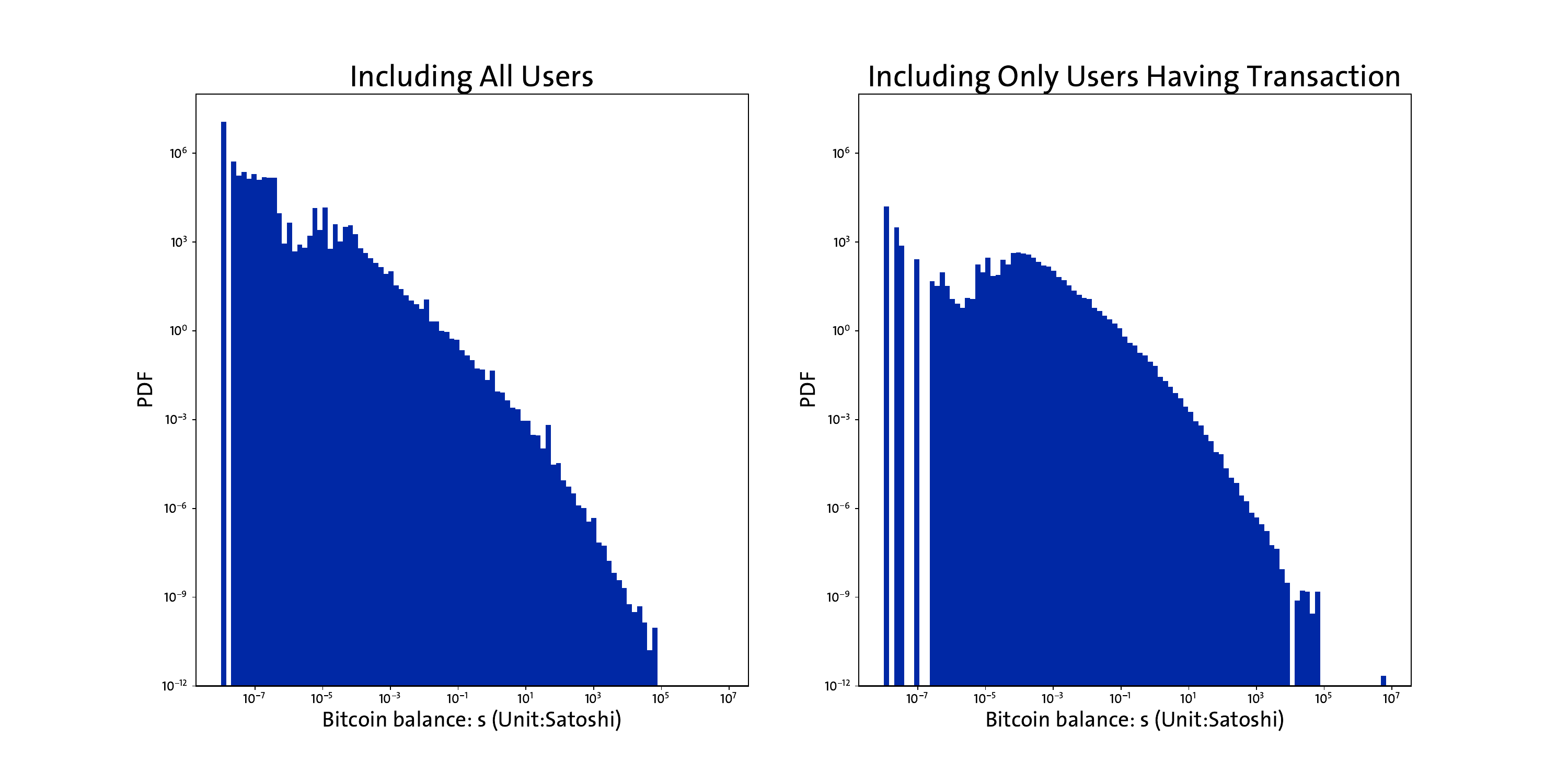}}
\caption{Bitcoin balance distribution on 2016-01-23 (unit: satoshi, 1 bitcoin=$10^{8}$ satoshi). The left panel depicts the probability distribution function of all users (groups A and B), and the right panel depicts only users whose balances are positive and who have transactions (group A) during the next period $dt$ starting from 2016-01-23 (it is 28 days in the right panel).}
\label{fig1}
\end{figure}

We chose the bitcoin balance data on 2016-01-23 because the bitcoin transaction network was more mature and relatively more stationary at that time compared with the earlier date. The log-log scale histogram in Fig. \ref{fig1} indicates that the users' bitcoin balance is a heavy-tail distribution. 

When it comes to the balance distribution, we need to distinguish two kinds of users. The first kind of users are those whose balances on 2016-01-23 are positive and who have transactions during the next period $dt$ (it is 28 days in the right panel), which is called user group A. The other kinds of users are those whose balances on 2016-01-23 are positive, but who do not have transactions during the next period $dt$, which is called user group B. This differentiation is important and necessary, otherwise, data from users who have not transacted for a long time will affect the accuracy of our analysis, for example, a dead bitcoin address.

The left panel in Fig. \ref{fig1} depicts the probability distribution function (pdf) with both kinds of users' balance data. The right panel in Fig. \ref{fig1} depicts the pdf with the balance data of only those (group B) whose balances on 2016-01-23 are positive and who have transactions during the next period $dt$. 

\begin{figure}[htbp]
\centerline{\includegraphics[width = 0.5\textwidth]{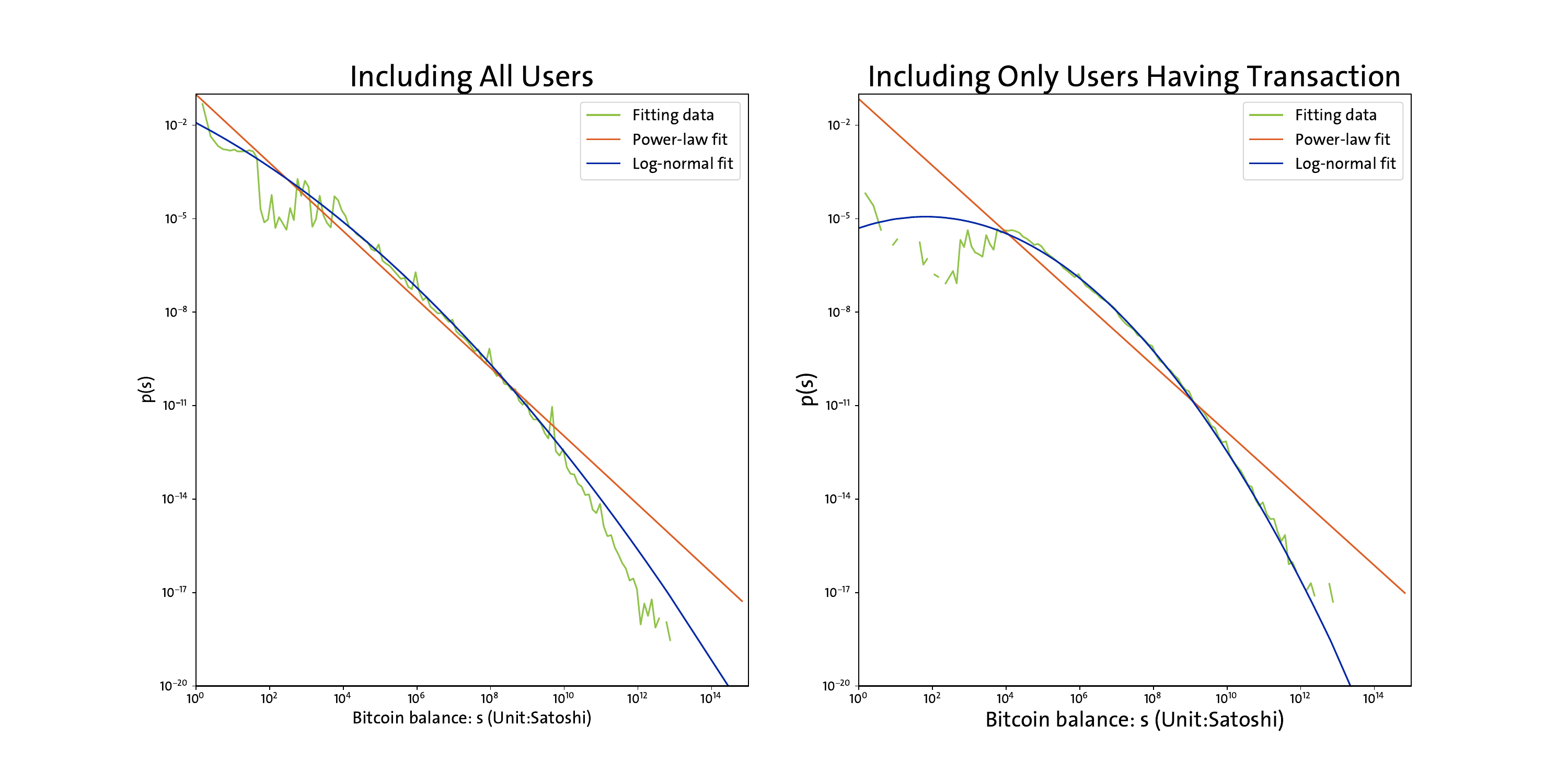}}
\caption{Bitcoin balance fitting (unit: satoshi). The left panel depicts the fitting results with the data from all users on 2016-01-23 (groups A and B), and the right panel depicts the fitting result with data from those users whose balances are positive and who have transactions (group A) during the next period $dt$ starting from 2016-01-23 (it is 28 days in the right panel).}
\label{fig2}
\end{figure}

\begin{figure}[htbp]
\centerline{\includegraphics[width = 0.6\textwidth]{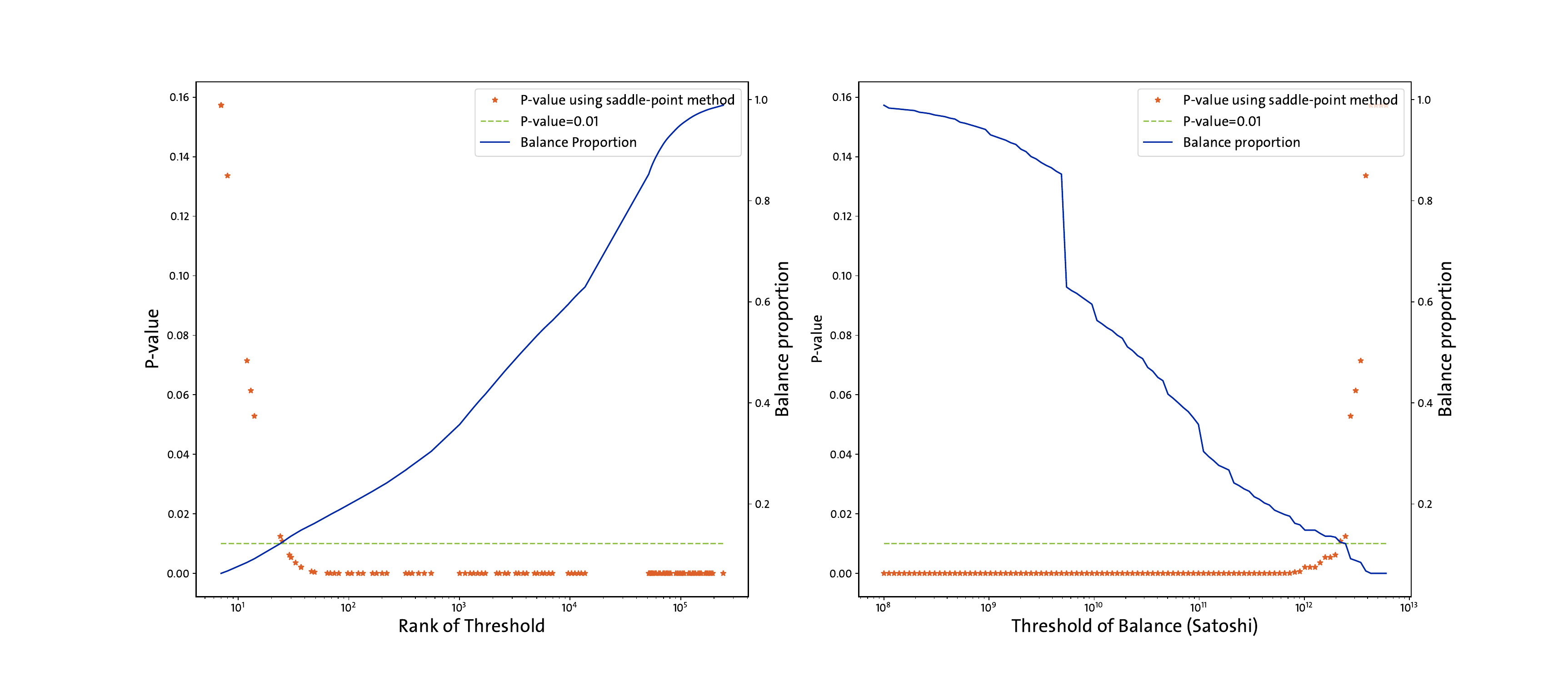}}
\caption{UMPU Wilks test results. The null hypothesis for this test is that the data is distributed as a power-law, and its alternative hypothesis is that the data is distributed as log-normal. The left panel is the p-value versus the rank of the chosen threshold of bitcoin balance on 2016-01-23. The smaller p-value, the more certain that the null hypothesis is refused. After the threshold is chosen, all bitcoin balance value above this threshold is sorted inversely, namely the largest value of bitcoin balance rank one, then the second largest value of bitcoin rank two. The ranking proceeds until the chosen threshold. The right panel is the p-value versus the threshold of bitcoin balance.}
\label{fig_p_value}
\end{figure}

Then applying the python power-law package developed by \cite{b3}, we fitted the bitcoin balance data on 2016-01-23 to the power-law and log-normal distribution and found that the log-normal distribution fits the data better. 
Fig. \ref{fig2} and statistical test in Fig. \ref{fig_p_value} also confirmed that the log-normal distribution is better than the power-law in fitting the balance distribution data. 

We also compared the fitting results between the log-normal distribution and power-law distribution by gradually increasing the minimum fitting data. The first minimum fitting data we choose is $10^{-8}$ bitcoin, and the increasing step is 1 bitcoin. That means that the minimum fitting data in the second time is 1 bitcoin, then 2 bitcoins in the third time, and so on. The result shows that the log-normal function is always a better fitting than the power law in most cases.  
Even if the data is generated by power-law distribution, we can't still refuse the hypothesis that the data is from a log-normal distribution only if its variance is huge just by fitting the data using the package developed by \cite{b3}. So, it is not enough to get the conclusion that our empirical data comes from power-law distribution or log-normal distribution only by this statistic package.

The uniformly most powerful unbiased (UMPU) Wilks test as suggested by \cite{b9} can be used to distinguish power-law distribution and log-normal distribution. This method comes from the idea that exponentiality can be tested against normal distribution \cite{b10}\cite{b11} using the saddle point approximation method and the idea that power-law distribution and log-normal distribution can be transferred to exponential distribution and normal distribution after taking log calculation to bitcoin balance data, respectively. The null hypothesis for this test is that the data is distributed as a power-law, and its alternative hypothesis is that the data is distributed as log-normal. The test is performed as follows: Firstly, we choose a threshold for the bitcoin balance; secondly, the UPMU Wilks test is performed for bitcoin balance whose value is larger than the threshold by computing the p-value. Though the Monte Carlo method can also be used to calculate the p-value, it is very time-consuming here because we have millions of data. As shown in Fig. \ref{fig_p_value}, we can reject the null hypothesis and accept the alternative hypothesis in almost all regions of bitcoin balance except regions that include only tens of the largest value of bitcoin balance. However, the proportion of tens of the largest value of bitcoin to the total number of bitcoins in our specific time-point is less than 5\%.

\subsection{Data Exploration}
Now that we know that the balance distribution may be log-normal, a natural question is what is the mechanism behind the transaction that leads to the log-normal distribution? 

Gibrat’s proportional growth law is an important tool in explaining the forming mechanism of power-law distribution if we change Gibrat’s proportional growth equation a bit \cite{b3}, and especially in Zipf's distribution when taking other mechanism into consideration, such as birth process and death process. However, we also understand that the probability density function (pdf) of the solution of standard Gibrat's proportional growth is log-normal distribution which is better than the power-law distribution in fitting our bitcoin balance data. At the same time, our test confirms that the bitcoin balance distribution function is fitted well by log-normal distribution. We also checked the distribution of bitcoin balance on 2019-01-19 which is almost three years later than 2016-01-23, and we find that the distribution is also log-normal. Can Gibratt's proportional growth law be the mechanism to explain our data? To answer this question, we need to investigate our data first and then check whether the exponent in Gibrat's proportional growth equation is one or neither.

\begin{figure}[htbp]
\centerline{\includegraphics[width = 0.5\textwidth]{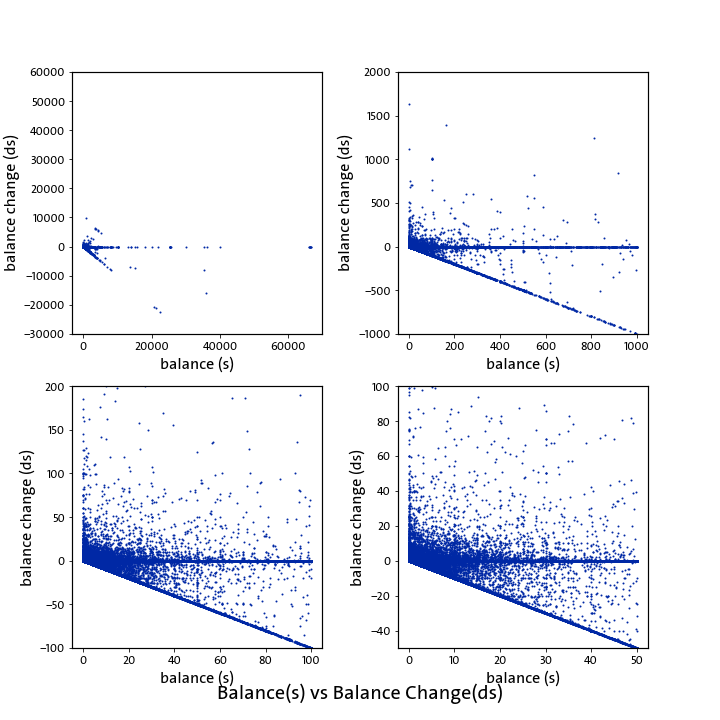}}
\caption{Bitcoin balance and balance change on 2016-01-23, the time interval (dt) is half a year.}
\label{fig_4}
\end{figure}

We first depict the scatter plot of bitcoin balance data ($s$) versus bitcoin balance change ($ds$) data to get a comprehensive impression of the holistic data distribution. Fig. \ref{fig_4} is the scatter plot of bitcoin balance versus bitcoin balance change within a half year. Four sub-scatter plots with different scales are depicted so that it is easy for us to look closely into the data.

There exist different clusters in the data. Hopkin statistics (smaller than $10^{-4}$ in the condition of 100 random samples) test (the null hypothesis is that there is only one cluster in the data; the alternative hypothesis is that there is more than one cluster in the data) also confirms the existence of multiple clusters. There are lots of methods for clustering data, such as distance-based K-means method, and probability-based Gaussian Mixture models. Both methods work well if data points are spared in a circle shape. Gaussian Mixture models also assume that it is Gaussian distributed in all dimensions of data points which is not the case in our data. Based on this analysis, to the best of our knowledge, there seem no methods that can be used to directly better cluster bitcoin users. 

However, as shown in Fig. \ref{fig_4}, there are three straight lines, the vertical line, the horizontal line, and the diagonal line that correspond to different situations. The vertical line corresponds to those users who don’t own or own only a small number of bitcoins on 2016-01-23 but got lots of bitcoins by trading before 2016-02-20 (28 days later after 2016-01-23). The horizontal line corresponds to those users who own bitcoins and the number of bitcoins didn’t change in the time interval between 2016-01-23 and 2016-02-20. The diagonal line corresponds to those users who owned bitcoins on 2016-01-23 but sold them all before 2016-02-20.


Based on our data exploration, we think that the data point on the horizontal line should be deleted because these corresponding users didn’t take part in trading activities in our specific time span.

\section{Mechanism Detection}
Now, we explore the mechanism behind bitcoin distribution.
As before, we still define $S_i$ as the cryptocurrency balance owned by the $i^{th}$ user. The change of cryptocurrency balance ($dS_i$) can be modeled as follows if they follow the Geometric Brownian Motion (GBM) mechanism:

\begin{equation}
    dS_i = {S_i}^\alpha \cdot \mu \cdot dt + {S_i}^\alpha \cdot \sigma \cdot dw_i
    \label{eq_ds_s}
\end{equation}
where $dS_i$ is the $i^{th}$ user' balance at the starting time; $dt$ is the time interval between the starting time and the ending time for measuring the balance change $dS_i$; $dS_i$ is the balance change of the $i^{th}$ user during $dt$; $w_i$ is the Brownian Motion, $\mu$ and $\sigma$ is the drift and volatility, respectively. $\alpha$ is the exponent we will focus on.
We can get the following equation by taking the expectation and variance on both sides of the equation \ref{eq_ds_s}:

\begin{equation}
\begin{cases}
    E(dS_i) = {S_i}^\alpha \cdot \mu \cdot dt \\
    \sigma(dS_i) = {S_i}^\alpha \cdot \sigma \cdot \sqrt{dt}
\end{cases}
\label{eq_2}
\end{equation}

We can plot the equation \ref{eq_2} to explore the parameters $\alpha$, $\mu$, $\sigma$ and $dt$. The whole process includes three main steps:
\begin{itemize}
    \item At first, the range of bitcoin balance is split as $n$ (for example, $n=300$) consecutive bins (with constant size or size that increases exponentially);
	\item Then, we classify the bitcoin balance data ($dS_i$) and corresponding bitcoin balance change data ($dS_i$) according to bins that we choose. After classifying, we delete those bins where the number of data ($dS_i$) is less than 50 (can be other numbers, like 100) and the corresponding bitcoin balance change data ($dS_i$). 
    \item At last, we calculate the average and standard deviation of $dS_i$ in each bin.
\end{itemize}

\begin{figure}[htbp]
\centerline{\includegraphics[width = 0.25\textwidth]{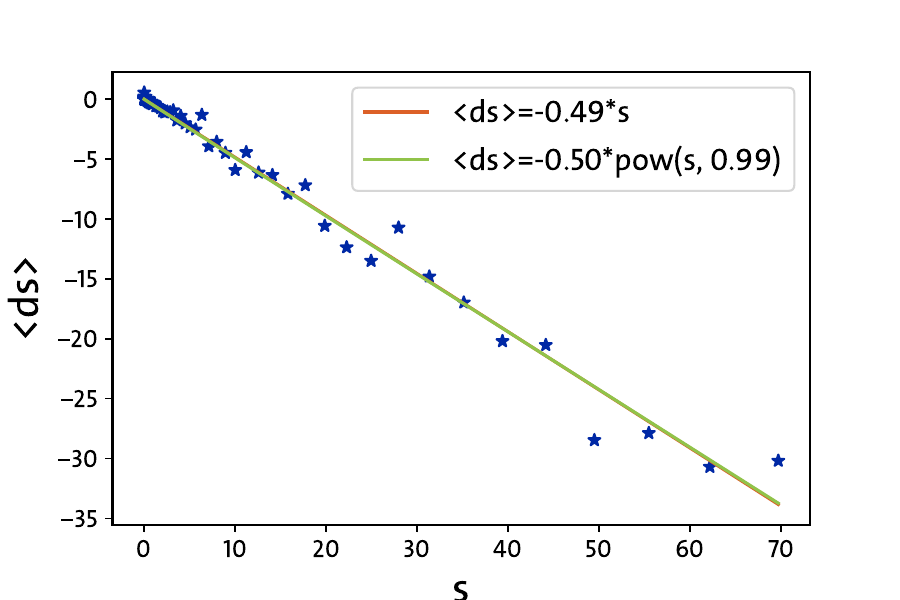}
\includegraphics[width = 0.25\textwidth]{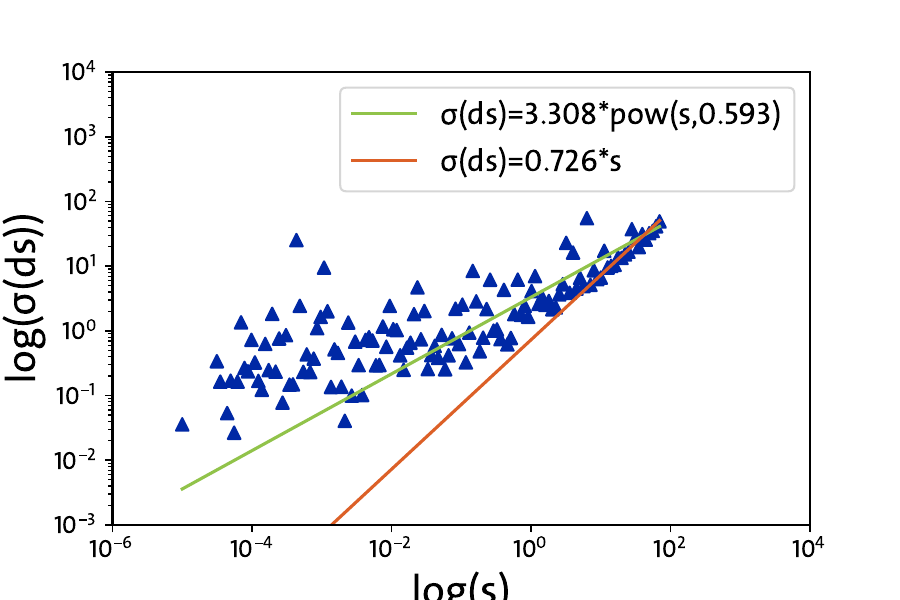}}
\caption{The left panel is bitcoin balance versus the average of bitcoin balance change; the right panel is bitcoin balance versus standard variation of bitcoin balance change.}
\label{fig_reg}
\end{figure}

Because the bitcoin balance distribution is scale-free, there are no data or only a few data points in lots of bins that correspond to large bitcoin balances, and most data are located in bins that correspond to several small balances. So, we think it would be a good choice to apply exponential bins and we got 167 data points which were shown in Fig. \ref{fig_reg}.

We calculated the fitting results based on equation \ref{eq_2} first. Because there are both negative and positive values in the average of a bitcoin balance change, it is not possible to use a log-log scale coordinate system to show the fitting equation \ref{eq_2}. So, we still use a constant scale coordinate system to show our data and fitting results. As shown in the left panel of Fig. \ref{fig_reg}, the red straight line corresponds to the case of proportional growth (exponent $\alpha$ is set to 1 in equation \ref{eq_2}), we only need to calculate the value of  $\mu \cdot dt$. The green line corresponds to the case that both exponent $\alpha$ and $\mu \cdot dt$ were calculated by fitting.  By comparing visually and making regressions, it seems that the exponent $\alpha$ is 1 in equation \ref{eq_2} can be accepted.

In the right panel of Fig. \ref{fig_reg}, the relationship between $\sigma (dS_i)$ and $S_i$ is shown. The red line corresponds to the case $\alpha=1$ by fitting the model $\sigma(dS_i) = {S_i} \cdot \sigma \cdot \sqrt{dt}$. 
The green line corresponds to the case in which we calculated the exponent $\alpha$ by making a regression $\sigma(dS_i) = {S_i}^\alpha \cdot \sigma \cdot \sqrt{dt}$ and $\alpha$ is 0.739 by fitting. The exponent value we get from the first equation in \ref{eq_2} is very different from the exponent value we get by fitting the second equation in \ref{eq_2}. Does this result denote that the exponent $\alpha$ in the volatility term is different from the exponent $\alpha$ in the drift term?

Because the absolute value of bitcoin balance change varies a lot for different users, we turn to research the ratio of balance change to balance $\frac{dS_i}{S_i}$. We get the following equation \ref{eq_3} by dividing $S_i$ in both sides of equation \ref{eq_2}: 

\begin{equation}
\begin{cases}
    E(\frac{dS_i}{S_i}) = {S_i}^{\alpha-1} \cdot \mu \cdot dt \\
    \sigma(\frac{dS_i}{S_i}) = {S_i}^{\alpha-1} \cdot \sigma \cdot \sqrt{dt}
\end{cases}
\label{eq_3}
\end{equation}

Based on equation \ref{eq_3}, there is no relationship between $E(\frac{dS_i}{S_i})$, $\sigma(\frac{dS_i}{S_i})$ and $S_i $ if $\alpha=1$. The only difference between our current calculation and previous ones is that we need to calculate the average and standard variance of $\frac{dS_i}{S_i}$  in each bin, now, but not $dS_i$.

\begin{figure}[htbp]
\centerline{\includegraphics[width = 0.6\textwidth]{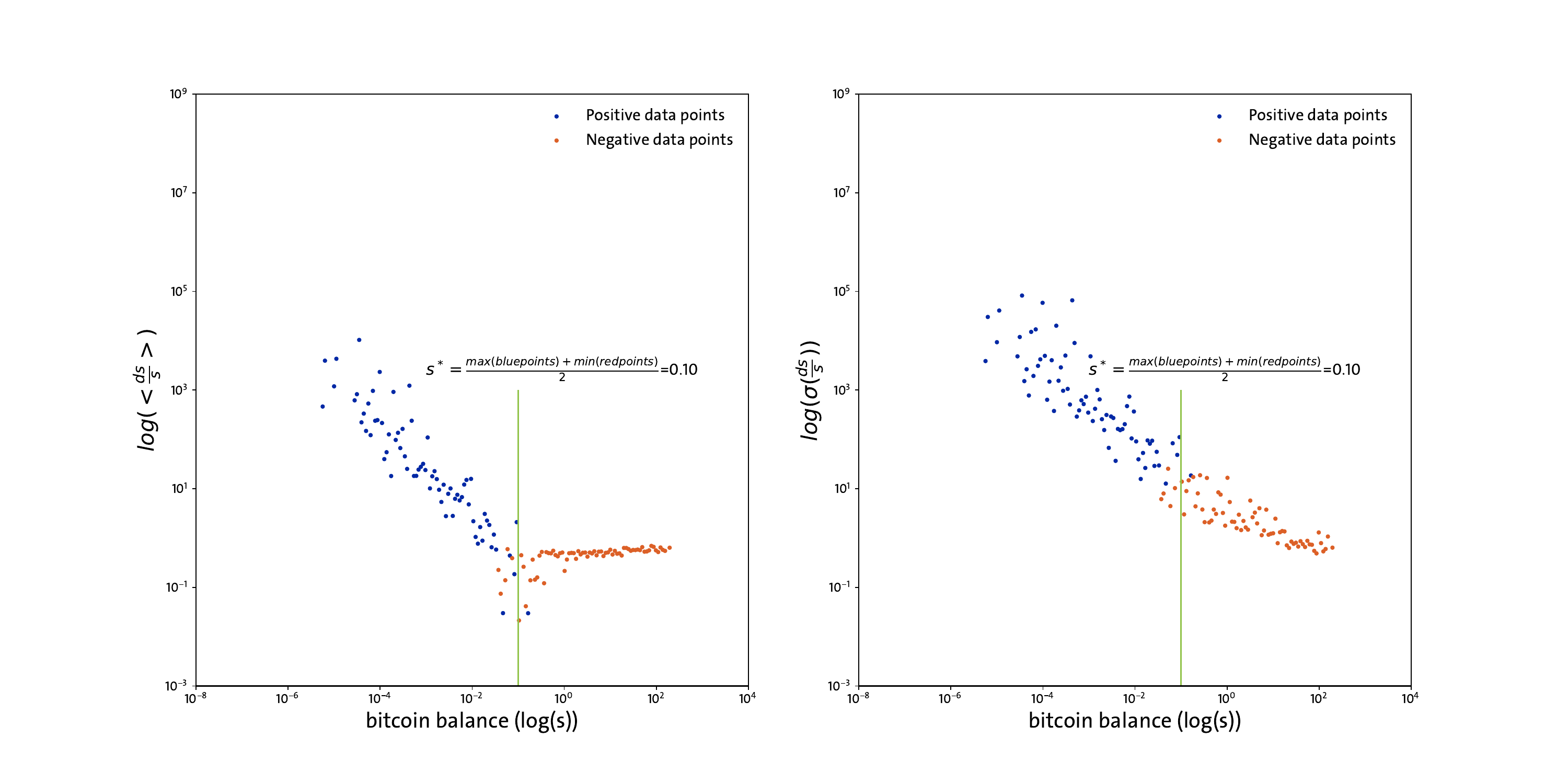}
}
\caption{The left panel is $ E(\frac{dS_i}{S_i})$ versus balance $S$, the right panel is the standard variance $\sigma(\frac{dS_i}{S_i})$ versus balance $S$. The coordinate is a log-log scale. The blue points correspond to these users whose $ E(\frac{dS_i}{S_i})$  is positive and red points correspond to those users whose $ E(\frac{dS_i}{S_i})$ is negative. Because the log function can’t be applied to a negative value, every negative value (red points) of $ E(\frac{dS_i}{S_i})$ needs to be multiplied by minus one to plot in a log-log scale. The green line corresponds to the average of the largest balance value of those blue points and the smallest balance value of those red points.
Note: The starting time is 2016-01-23, and the time interval ($\Delta t$) is 112 days (almost four months).
}
\label{fig_dss_av_std}
\end{figure}

As Fig. \ref{fig_dss_av_std} shows, surprisingly, there are two different modes (blue points and red points) for the users' balance changes. For users whose bitcoin balance is less than a specific value (blue line, we also called them \textbf{poor bitcoin users}), the average of $ \frac{dS_i}{S_i}$  is positive, namely $\mu>0$ in equation \ref{eq_3}. The left panel of Fig. \ref{fig_dss_av_std} shows that there is a linear relationship between $ E(\frac{dS_i}{S_i})$ and balance $S$, and the slope of this linear line is negative which means that  $\alpha<1$ for those blue points.  
The right panel of Fig. \ref{fig_dss_av_std} shows that there is also a negatively correlated relationship between $\sigma(\frac{dS_i}{S_i})$  and $S_i$, which denotes again $\alpha<1$ for those blue points.
By contrast, for users whose bitcoin balance is larger than a specific value (red points, we also called them \textbf{wealthy bitcoin users}), the average of $ \frac{dS_i}{S_i}$  ($ E(\frac{dS_i}{S_i})$) is negative, which means that $\mu<0$ in equation \ref{eq_3} for red points. The line in the left panel of Fig. \ref{fig_dss_av_std} is almost horizontal, but upward a bit actually, which means that $\alpha$ is larger than or close to one for red points. However, the right panel of Fig. \ref{fig_dss_av_std} shows that the linear line is not exactly horizontal for red points, which means that $\alpha<1$ for the volatility term. 
These analyses show that there exist two different balance growth models for Bitcoin users, which are as follows:

\begin{equation}
\begin{cases}
    dS_i = {S_i}^{\alpha_{<1}^{11}}\cdot \mu_{>0} \cdot dt + {S_i}^{\alpha_{<1}^{12}} \cdot \sigma dw_i \quad (S_i<S^*)
    \\
    dS_i = {S_i}^{\alpha_{>1}^{21}}\cdot \mu_{<0} \cdot dt + {S_i}^{\alpha_{<1}^{22}} \cdot \sigma dw_i \quad (S_i<S^*),
\end{cases}
\label{eq_4}
\end{equation}
where subscript $<1$, $=1$, $>0$, and $<0$ denote that the corresponding value is smaller than one, equal to one, larger than zero, and smaller than zero, respectively. 
For example, ${\alpha_{<1}^{11}}<1$, $\mu_{>0}>0$.
$S^*$ is a threshold value. That means that for users whose bitcoin balance value is smaller than $S^*$, their balances grow according to the first model of equation \ref{eq_4}. Because the corresponding exponent $\alpha<1$ and $\mu>0$, we can’t get an analytical solution for this model. So, we can’t calculate exactly how the bitcoin balance of these users will change on average with time. For users whose bitcoin balance value is larger than $S^*$, their balances grow according to the second model of equation \ref{eq_4}. 

\begin{figure}[!ht]
\begin{subfigure}{0.5\textwidth}
\centering
\includegraphics[width = 1\linewidth]{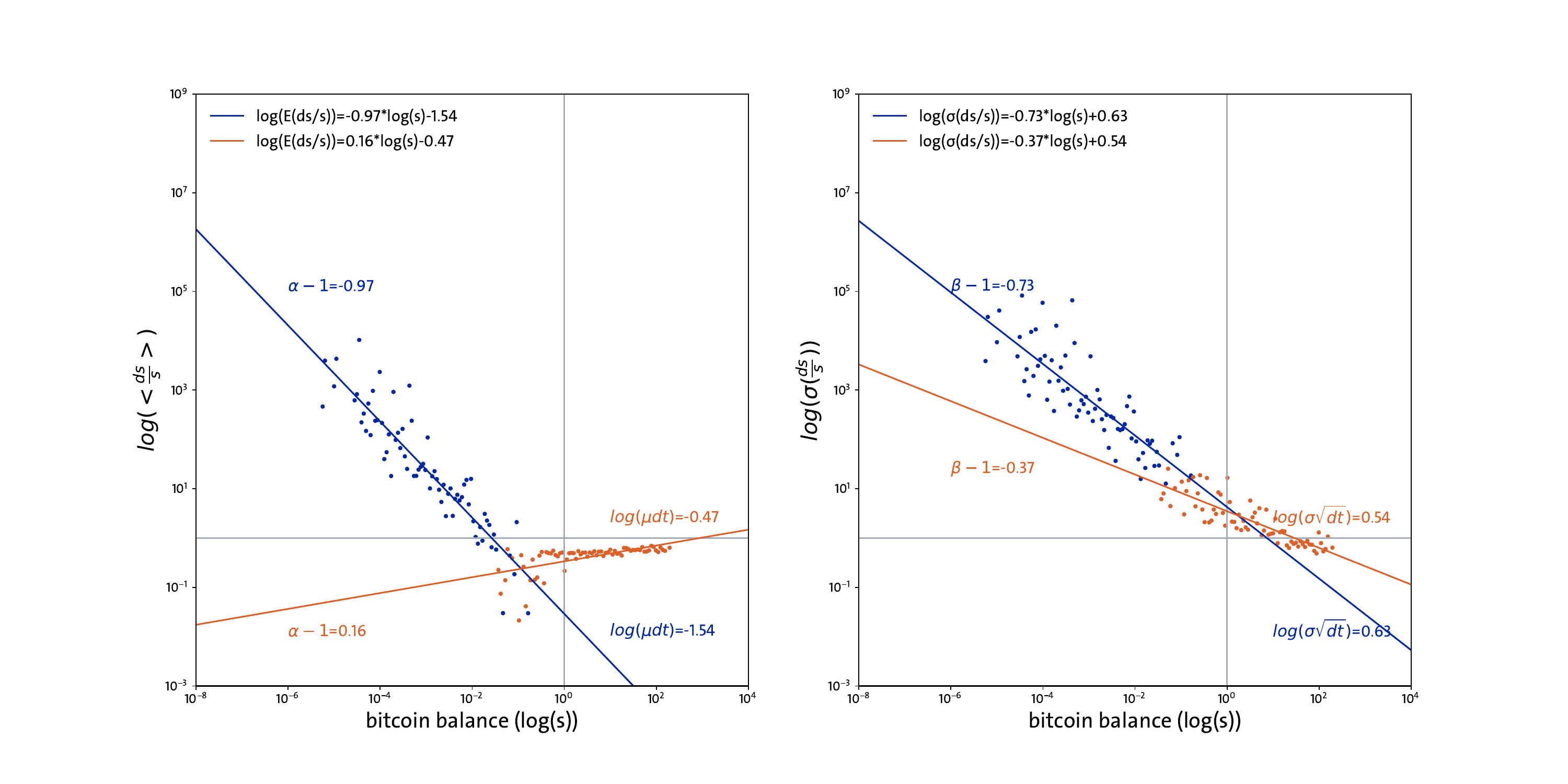}
\caption{Parameter fitting}
\label{1}
\end{subfigure}
\begin{subfigure}{0.5\textwidth}
\includegraphics[width = 1\linewidth]{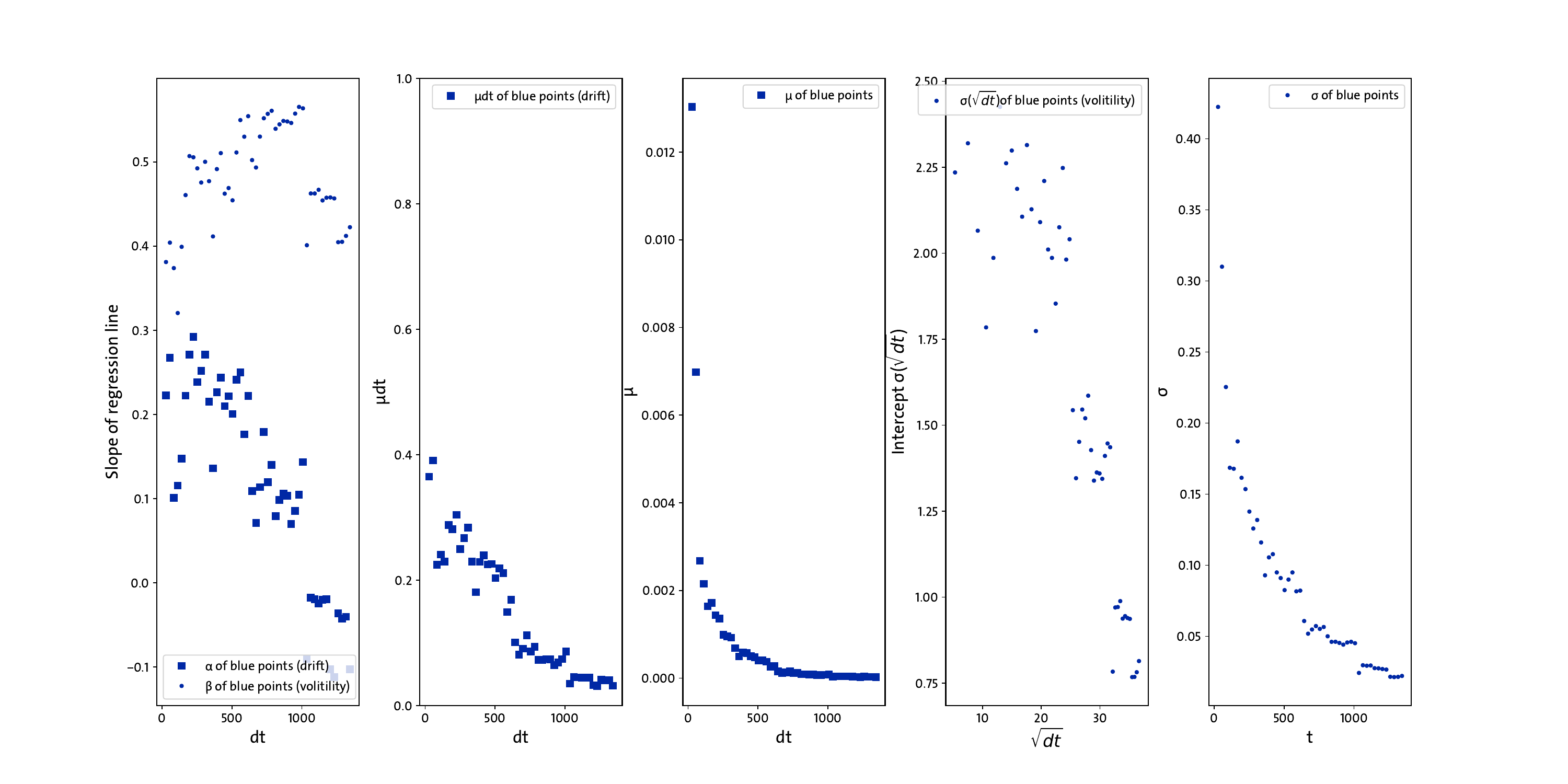}
\caption{Parameter with $dt$ of poor bitcoin users}
\label{2}
\end{subfigure}
\begin{subfigure}{0.5\textwidth}
\includegraphics[width = 1\linewidth]{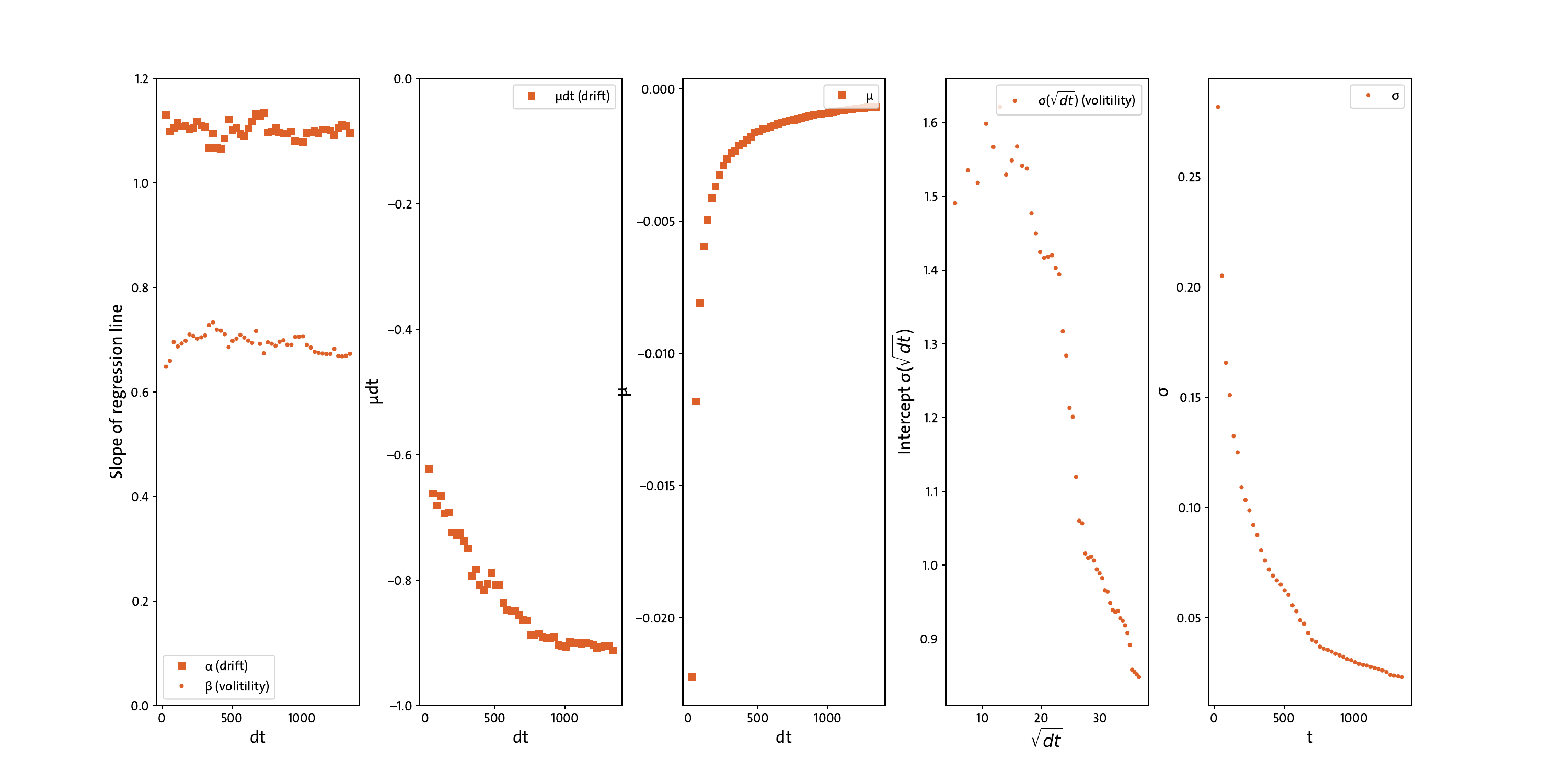}
\caption{Parameter with $dt$ of wealthy bitcoin users}
\label{3}
\end{subfigure}
\caption{Panel (a) shows how to calculate related parameters by regression.
Panel (b) and (c) depict the regressed parameters with time interval $dt$ increasing for poor and wealthy bitcoin users. The parameters include $\alpha$ in both drift and volatility term,  $\mu \cdot dt$, $\sigma \cdot \sqrt{dt}$. 
Note: The time interval ($dt$) changes from 1 month to 24 months. The unit of the x-axis in panels (b) and (c) is day.
}
\label{fig_parameter_fitting}
\end{figure}

We now research whether this type of growth model is stable by changing the time interval $dt$, by which we can also explore how bitcoin users behave with time. Our researching target includes the exponent $\alpha$, drift parameter $\mu$, and volatility parameter $\sigma$. 
For every time interval $dt$, we calculated the average ($ E(\frac{dS_i}{S_i})$) and the variance ($ \sigma(\frac{dS_i}{S_i})$) of $\frac{dS_i}{S_i}$  in each bin of bitcoin balance. Then, we got these target parameters ($\alpha$, $\mu$, $\sigma$) by making a linear regression between the average ($E(\frac{dS_i}{S_i})$), variance ($\sigma(\frac{dS_i}{S_i})$) and balance S, respectively, as shown in sub-Fig. \ref{1}.

As shown in sub-Fig. \ref{2}, for the first stochastic equation ($dS_i = {S_i}^{\alpha_{<1}^{11}}\cdot \mu_{>0} \cdot dt + {S_i}^{\alpha_{<1}^{12}} \cdot \sigma dw_i \quad(S_i<S^*)$) in equation \ref{eq_4}, the value of exponent $\alpha$ in both drift term and volatility term fluctuate a lot but are both less than 1.
The exponent $\alpha$ in the drift term is negatively correlated with the time interval ($dt$), while the exponent $\alpha$ in the volatility term seems constant despite of much fluctuation. 
The second figure in sub-Fig. \ref{2} shows that the parameter $\mu dt$ is a monotonically decreased function with time interval ($dt$) but it tends to zero in the last. It means that poor bitcoin users initially buy very few bitcoins will buy more in the next time interval $\Delta t$ and then sell all bitcoins gradually in the future.
The fourth and fifth figures of sub-Fig. \ref{2} shows that  $\sigma \cdot \sqrt{dt}$ fluctuates a lot with time interval ($dt$) but is negatively correlated with time interval ($dt$), and $\sigma$ decreases with $dt$. 
This means that poor bitcoin users tend to sell all their coins with $dt$ increasing for sure.
By analyzing, the exact formula of the equation should be:

\begin{equation}
    dS_i = {S_i}^{{\alpha(dt)^+}_{<1}^{11}}\cdot {\mu(dt)^-}_{>0} \cdot dt + {S_i}^{\alpha_{<1}^{12}} \cdot \sigma (dt)^-dw_i \quad(S_i<S^*);
\label{eq_5}
\end{equation}
where ${{\alpha(dt)^+}_{<1}^{11}}$ denotes that exponent $\alpha$ in drift term is a monotonically increased function with time interval ($dt$) and smaller than 1; ${\mu(dt)^-}_{>0}$ denotes that $\mu$ is a monotonically decreased function with time interval ($dt$) and larger than zero; ${\alpha_{<1}^{12}}$ denotes that exponent $\alpha$ in volatility term is constant and smaller than 1; $\sigma (dt)^-$ denotes that $\sigma$ is a monotonically decreased function with time interval ($dt$).


Now, we focus on analyzing the second formula of equation \ref{eq_4} ($dS_i = {S_i}^{\alpha_{>1}^{21}}\cdot \mu_{<0} \cdot dt + {S_i}^{\alpha_{<1}^{22}} \cdot \sigma dw_i \quad(S_i>S^*)$), where $\mu<0$. 
As shown in the first figure of sub-Fig. \ref{3}, the value of exponent $\alpha$ in both drift term and volatility term is nearly constant with time interval. However, the exponent $\alpha$ in the drift term is larger than one, and it is smaller than one in the volatility term. The third figure of sub-Fig. \ref{3} shows that the drift term parameter $\mu \cdot dt$ is negative and decreases with $dt$, which means that these users who own lots of bitcoins have the trend to sell their bitcoins with time flying.

The fourth figure of sub-Fig. \ref{3} denotes that  $\sigma \cdot \sqrt{dt}$ decreases with $\sqrt{dt}$, so the parameter $\sigma$ in volatility term of the second equation in equation \ref{eq_4} should be some monotonically decreased function of time interval $dt$, which means that wealthy bitcoin users will sell their bitcoins for sure over $dt$.

By analyzing, the exact formula of the equation it should be:

\begin{equation}
    dS_i = {S_i}^{\alpha_{>1}^{21}}\cdot \mu_{<0} \cdot dt + {S_i}^{\alpha_{<1}^{22}} \cdot \sigma(dt)^- dw_i \quad (S_i>S^*),
\label{eq_6}
\end{equation}
where $\sigma(dt)^-$ denotes that $\sigma$ is a monotonically decreased function with time interval $dt$. Other parameters ($\alpha$, $\mu$) in this equation are constant.
The model will be:

\begin{equation}
\begin{cases}
    dS_i = {S_i}^{{\alpha(dt)^+}_{<1}^{11}}\cdot {\mu(dt)^-}_{>0} \cdot dt + {S_i}^{\alpha_{<1}^{12}} \cdot \sigma (dt)^-dw_i \quad (S_i<S^*)
    \\
    dS_i = {S_i}^{\alpha_{>1}^{21}}\cdot \mu_{<0} \cdot dt + {S_i}^{\alpha_{<1}^{22}} \cdot \sigma(dt)^- dw_i \quad (S_i>S^*).
\end{cases}
\label{eq_7}
\end{equation}

\section{Summary and Discussion}
In this paper, we explore the transaction patterns of bitcoin users. Firstly, we explored users' balance distribution and found that the log-normal distribution function can better fit the balance. Secondly, we explored whether bitcoin users' transaction behavior follows Gibrat’s proportional growth rule and found that their transaction behaviors didn't follow Gibrat’s proportional growth rule. By extending related analysis, we find that there exist two kinds of bitcoin users: wealthy users who own plenty of bitcoins, in the beginning, tend to sell their bitcoins; poor users who have few bitcoins, in the beginning, tend to buy a bit in next period and then sell all their bitcoins again in the future.

By analyzing the balance data for wealthy users, we found that the exponent of $S$ in the drift term is almost constant and slightly larger than one, and the exponent of $S$ in the volatility term is also almost constant and slightly smaller than one, which was shown in the second equation in equation \ref{eq_7}.

The UTXO-based blockchain records each coin's flow history and provides us the chance to research human economic behaviors. The research on the patterns of users' transaction behaviors on UTXO-based blockchain is still in its early stages and deserves more attention in the future. This paper provides a good starting point in this direction and the research results may also be applicable in other traditional fields.

\vspace{12pt}

\end{document}